\documentclass[10pt,aps,prb,twocolumn,superscriptaddress,tightenlines,floatfix]{revtex4-2}
\usepackage{graphicx} % Required for inserting images
\usepackage{bm}
\usepackage{amsmath}
\usepackage{amssymb}
\usepackage{physics}
\usepackage{xcolor}
\usepackage{comment}
\usepackage{mathtools}
\usepackage{bbold}
\usepackage{enumitem}
\usepackage{ulem}
\renewcommand{\S}{\bm{S}}
\usepackage{nicematrix}
\newcommand{\KR}[1]{\textcolor{red}{\textbf{#1}}}

\newcommand{\dg}{^{\dagger}}
\newcommand{\bk}{{\bf k}}

\usepackage[pdfencoding=auto, psdextra]{hyperref}

\hypersetup{
    colorlinks = true,
    linkcolor = blue
}

\begin{document}

\title{The fate of \texorpdfstring{$s_{\pm}$}{s+-} superconductivity in the bilayer nickelate \texorpdfstring{La$_3$Ni$_2$O$_7$}{La3Ni2O7}}
\author{Kaustubh Roy}
\affiliation{Department of Physics and Astronomy, Rutgers University}
\author{Piers Coleman}
\affiliation{Department of Physics and Astronomy, Rutgers University}
\affiliation{Department of Physics, Royal Holloway, University of London}
\date{\today}

\begin{abstract}
    Motivated by the recent measurement of high-temperature superconductivity in the bilayer nickelate system La$_3$Ni$_2$O$_7$ in the bulk \cite{SunWang2023} as well as thin film samples, we seek to understand the role of the interlayer coupling in this system in dictating the pairing symmetry. Many proposals posit the existence of an $s_\pm$ superconducting phase in close competition with the mroe conventional cuprate-like $d$-wave state in both weak- and strong-coupling theories. However, at least in the latter, the effects of Coulomb repulsion and the double occupancy constraint have been largely neglected. In this manuscript, we propose a mean-field fractionalization treatment that captures the full $SU(2)$ particle-hole symmetry of the t-J model at half-filling, and show how a systematic treatment of this new symmetry leads to an added Coulomb pseudopotential constraint - one that is absent in the usual $U(1)$ treatment. We then apply this new fractionalization scheme to a simplified bilayer t-J model of La$_3$Ni$_2$O$_7$, showing how the Coulomb pseudopotential suppresses the existence of $s_\pm$ symmetry in favor of $d$-wave pairing.
\end{abstract}

\maketitle

\section{Introduction}
\begin{comment}
\textcolor{blue}{
\textbf{Bullet points from discussion:}
\begin{itemize}
    \item Think of the nickelates as a heavily doped cuprate
    \item Our theory is an effective $3d_{x^2 - y^2}$ dominated theory with bonding and antibonding orbitals
    \item Just like how the cuprates have a one-band effective theory of Zhang-Rice singlets, we assume this theory of nickelates does too.
\end{itemize}
}
\end{comment}

The discovery of high-temperature superconductivity in the cuprates is a landmark achievement of physics \cite{BednorzMuller1986,Lee2006,Keimer2015} which has since
spurred the community to consider other related oxides, such as the ruthenates~\cite{Maeno1994Sr2RuO4,Rice1995Sr2RuO4} and more recently, the nickelates. Anisomov and Rice~\cite{AnisimovRice1999} originally reasoned that nickel,  with one less electron than copper, might, in a strongly reducing  environment, be pushed towards the $d^8/d^9$ 
configuration of copper yielding a new family of high temperature superconductors.  The chemistry required to fulfill their proposal required three decades of further  development in solid-state synthesis, culminating in the discovery of superconductivity in the infinite layer nickelates~\cite{Li2019Nickelate}. 
Further advances in synthesis and epitaxy over the past decade now leave us at a point where nickelate transition temperatures are comparable with the cuprates, and the discovery of ambient pressure materials opens up the potential of experiments at the same level of detail as in the cuprates~\cite{Puphal2026Nickelates}. This makes it an ideal time to revisit the theory.

A variety  of theories have been put forth to explain how strong correlations between electrons drive pairing\cite{Lee2006,Scalapino2012}. A  central question that each of these theories must answer is how electrons can accommodate the effects of strong Coulomb repulsion?  In  Bardeen Cooper Schrieffer (BCS) theory, the attractive interaction mediated by phonon exchange avoids the Coulomb interaction through the effect of retardation \cite{MorelAnderson1962,Bauer2013}: in simple terms, the deformation of the lattice  produced by an electron remains long after it has passed.  However, in strongly interacting systems where the  bare interaction between the electrons exceeds their kinetic energy, the pairing fermions avoid the Coulomb interaction by developing nodes in the pair wavefunction: nodes cost kinetic energy, but they pay off by keeping the electrons apart and minimizing the Coulomb cost.  

%The strength of onsite Coulomb interaction is often characterized by a Hubbard-like term in the Hamiltonian:

%\begin{equation}
%    H_{\text{int}} = U \sum_i n_{i\uparrow} n_{i\downarrow}.
%\end{equation}
%where $n_{i\sigma} = c^\dagger_{i\sigma} c_{i\sigma}$ is the fermionic occupancy operator at site $i$, with spin index $\sigma$.

\begin{figure}
    \centering
    \includegraphics[width=0.6\linewidth]{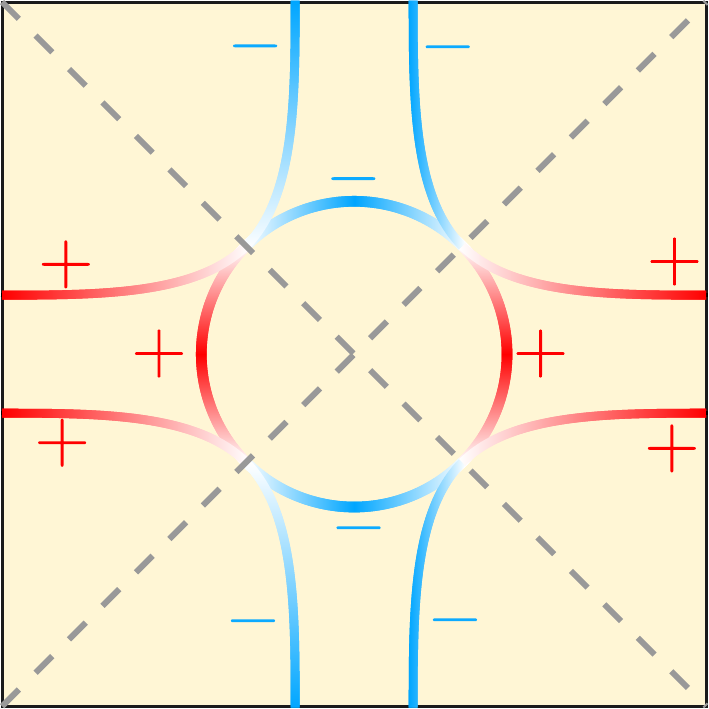}
    \caption{$d$-wave pairing order. The lines nodes running across the Brillouin zone (black dotted lines) break up the regions of changing signs of the pair gap function}
    \label{fig:dwave}
\end{figure}

In a weak-coupling treatment the development of nodes in the pair wavefunction can be seen from the zero-temperature gap equation,
\begin{equation}
    \Delta_{\bm k} = -\sum_{\bm k', |\epsilon_{\bk}|<\Lambda} V_{\bm k, \bm k'} \left(\frac{\Delta_{\bm k'}}{2 E_{\bm k'}}\right),
\end{equation}
where $\Lambda$ is a low energy cut-off. 
If we rewrite the gap equation in the form
\begin{equation}
    \Delta_{\bm k} = -\sum_{\bm k', |\epsilon_{\bk}|<\Lambda}
    {\rm sgn}(V_{\bm k, \bm k'}){\rm sgn}(\Delta_{, \bm k'})
    \left| V_{\bm k, \bm k'} \left(\frac{\Delta_{\bm k'}}{2 E_{\bm k'}}\right)\right|,
\end{equation}
then it is clear that if the interaction  is predominantly repulsive ($V_{\bm k, \bm k'}>0$), then there is no solution for the gap function $\Delta_{\bm k}$ that is positive everywhere on the Fermi surface. Moreover, to maximize the gap it is  advantageous that ${\rm sgn}(V_{\bm k, \bm k'}){\rm sgn}(\Delta_{\bm k'})<0$, i.e that the gap $\Delta_{\bf k}$ must become  negative in regions where $V_{{\bf k}, {\bf k}'}>0 $ is repulsive, giving rise to an anisotropic  gap with nodes. Famously, the gap $\Delta_{\bf k}$ in superfluid Helium-3  has  $p$-wave symmetry \cite{Leggett1975}, while that of cuprate superconductors has $d$-wave  symmetry (Fig. \ref{fig:dwave}). The presence of $d$-wave pairing in the cuprate superconductors has been confirmed by multiple experimental methods, such as corner-sharing SQUID experiments\cite{Wollman1993dWave}, flux quantization experiments\cite{Tsuei1994dWave}, angle-resolved photoemission spectroscopy (ARPES) \cite{DamascelliShen2003} and scanning tunnelling microscopy (STM) \cite{RennerFischer1995, HoffmanMcElroy2002, PanDavis2001}.

\begin{figure}[b]
    \centering
    \includegraphics[width=0.85\linewidth]{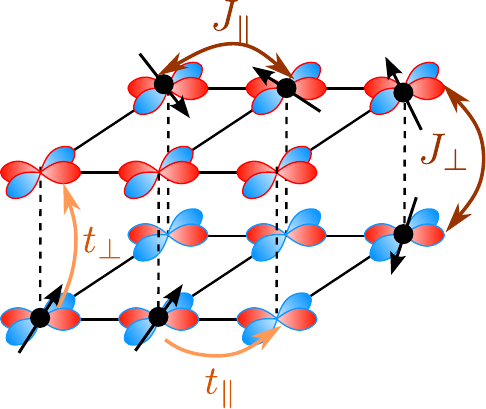}
    \caption{Bilayer t-J model of the bilayer nickelate La$_3$Ni$_2$O$_7$ adopted for our analysis. }
    \label{fig:placeholder}
\end{figure}

%The notable theories that have stood the test of time can be grouped into two categories. The former is the weak-coupling regime, where the superconductivity is mediated by spin- or charge fluctuations in a formulation akin to BCS. \KR{is this true??? Maybe I should rephrase this to say "strongly correlated superconductivity"} The latter is the strong-coupling regime, where antiferromagnetic superexchange interactions dominate the low-energy physics and the superconducting condensate can be thought of as emerging from a resonating valence-bond (RVB) ground state. 
While weak-coupling theory provides a qualitative  understanding of the development of gap anisotropy and the development of $d$-wave symmetry under the influence of antiferromagnetic interactions\cite{Scalapino1986dWave,Bickers1989Hubbard,MonthouxBalatskyPines1991}, in the cuprates the bare strength of interactions is comparable with their band-width, taking them out of the realm of weak-coupling.  One approach is to try to renormalize the physics from weak to strong coupling\cite{HalbothMetzner2000}.  The alternative attempts to incorporate strong coupling effects at the outset. A well-known strong-coupling  model for cuprate superconductors is
\textit{the $t$-$J$ model}\cite{ZhangRice1988}
\begin{equation} \label{eq:tJ}
    H = -t \sum_{\langle i,j\rangle} X_{\sigma0}^{(i)}X_{0\sigma}^{(j)} + J\sum_{\langle i,j\rangle} \bm S_i \cdot \bm S_j,
\end{equation}
%\KR{gutzwiller projection takes care of no double occupancy, but most studies of the t-J model like the ones below did not care to be careful about this constraint.}
%\KR{A list of the citations needed for the comparison}\cite{OhZhang2023, LuoYao2024,XueWang2024, LuWu2024a, LuWu2024b, PanWu2024, DuanSi2025}
which describes electrons hopping with amplitude $-t$ between neighboring sites on a square lattice,  with an additional antiferromagnetic coupling  $J$ between two neigboring electrons. The antiferromagnetic interaction arises from integrating out the states of double occupancy from the Hubbard model, giving rise to a super-exchange coupling  between sites of magnitude $J \sim t^2/U$, where $U$ is the onsite interaction energy of the Hubbard model. The $t$-$J$ model is defined on the subspace that excludes double occupancy on each site, spanned by the Hubbard operators $X_{\alpha\beta} = \ket{\alpha}\bra{\beta}$, where $\ket{\alpha},\ket\beta \in \{\ket0, \ket\uparrow, \ket\downarrow\}$. These non-canonical operators are neither boson nor fermion, but instead obey the Hubbard superalgebra ($[X_{\alpha\beta},X_{\gamma\delta}]_\pm = \delta_{\beta\gamma} X_{\alpha\delta} \pm \delta_{\alpha\delta} X_{\gamma\beta}$)\cite{Hubbard1965}.

%\KR{Say something about the gauge constraint}
\subsection{Contrasting the nickelates and cuprates}

\begin{figure}
    \centering
    \includegraphics[width=0.6\linewidth]{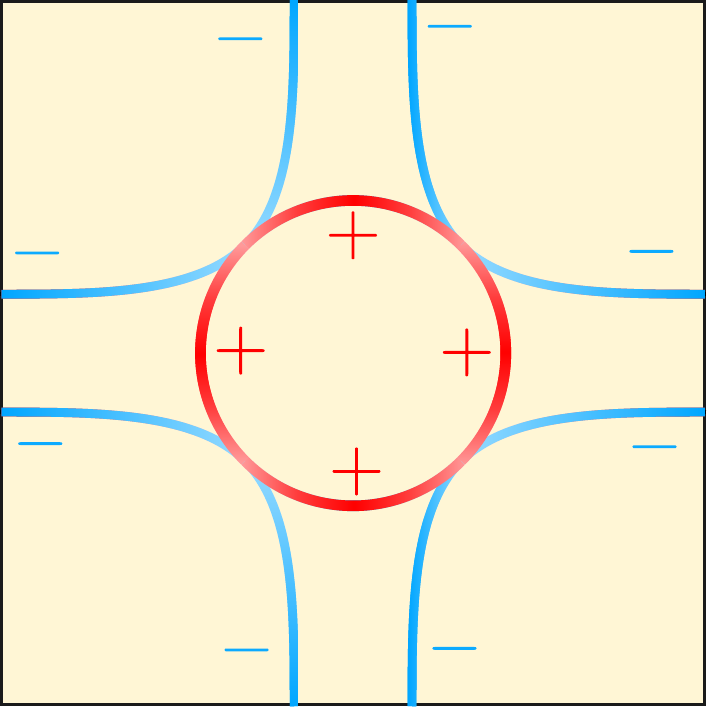}
    \caption{$s_\pm$-wave pairing order. The extended-$s$ pairing is superimposed with an extra pairing gap $\Delta_z$ that changes sign between Fermi surfaces.}
    \label{fig:spm_wave}
\end{figure}

In both the nickelates and the cuprates, the d-electrons are subject to a cubic crystal field from the neighboring oxygen atoms, leading to the  $e_g$ shell ($3d_{x^2 - y^2}, 3d_{z^2}$) being higher in energy. The physics of the both classes of material is dominated by the  $e_g$ states, while the  $t_{2g}$ orbitals are fully filled even at moderate hole doping and do not participate in the physics.  
%In spite of their similarities, early theoretical calculations \cite{AnisimovRice1999} in the nickelates comparable to the cuprates. 
%A further twenty years
Until only a few years ago, the superconducting $T_c$s in the nickelates were limited below 30 K. Amongst these early materials   Nd$_{0.82}$Sr$_{0.18}$NiO$_2$ thin films \cite{WangCheng2022}) achieved the highest $T_c$s.  In these materials, the Ni hosts an oxidation of +1, endowing it with an outer shell configuration identical to the cuprates ($3d^9$). In 2023, the Ruddlesden-Popper nickelate La$_3$Ni$_2$O$_7$ was synthesized with a $T_c$ of 80 K \cite{SunWang2023}. This system is markedly different from the cuprates - the Ni cation has an oxidation of +2.5, corresponding to an outer shell configuration of $3d^{7.5}$. Unlike the cuprates, which can be modeled with a single Fermi surface, La$_3$Ni$_2$O$_7$ hosts multiple Fermi surfaces, leading many groups to propose the possibility of $s_\pm$ pairing - an isotropic superconducting gap that switches sign on different Fermi pockets, as illustrated in   Fig. \ref{fig:spm_wave}. Similar proposals have been explored for the iron pnictides.

Theories of pairing in La$_3$Ni$_2$O$_7$ range from weak-coupling treatments of spin- and charge-fluctuation-mediated pairing \cite{GuHu2025,YangWang2023} to strong-coupling, RVB-based descriptions \cite{OhZhang2023,LuoYao2024,XueWang2024,LuWu2024a,LuWu2024b,PanWu2024,DuanSi2025}. Recent work has sharpened the competition between $s_\pm$ and $d$-wave pairing. A microscopic two-orbital $t$--$J$ derivation identifies exchange processes beyond the simplest single-orbital model \cite{KanekoKakoiKuroki2025}, while cluster quantum Monte Carlo calculations find a leading interlayer $s_\pm$ instability in a two-orbital Hubbard--Hund model \cite{MaierDoak2026}. Constrained-path Monte Carlo instead identifies a crossover between $d$-wave and $s_\pm$ pairing controlled by Hund's coupling and crystal-field splitting \cite{XiongCaiMa2026}. Other multiorbital calculations find a strong sensitivity to spin-fluctuation character and even possible odd-frequency pairing \cite{Gao2025}. For ambient-pressure films, weak- and strong-coupling calculations find interlayer $s_\pm$ pairing even without a $\gamma$ Fermi-surface pocket \cite{ShaoLu2026}. Finally, functional-renormalization-group calculations show that nonlocal Coulomb interactions can either suppress or promote interlayer $s_\pm$ pairing, depending on their orbital structure \cite{ZhanLeWu2026}.

These studies incorporate electronic correlations in substantially more detailed orbital models, but address a different question from the one posed here. We deliberately retain only a minimal bilayer $d_{x^2-y^2}$ sector, with the effects of the other orbitals encoded in effective couplings, in order to isolate the consequences of enforcing the local no-double-occupancy constraint on the pairing saddle point. In particular, a sign reversal between Fermi sheets does not by itself eliminate the on-site anomalous amplitude, whereas a symmetry-pure $d_{x^2-y^2}$ gap has a vanishing on-site projection in the tetragonal limit. Our $SU(2)$ construction enforces this constraint self-consistently, allowing us to quantify its effect on the competition between interlayer $s_\pm$ and intralayer $d$-wave pairing. Thus our result is not that multiorbital $s_\pm$ pairing is impossible, but that it must satisfy an additional local condition whose impact can be exposed most clearly in a minimal model.

The distinction between the two materials thus motivates the question: how can one compare interlayer $s_\pm$ and intralayer $d$-wave pairing on equal footing while excluding forbidden on-site pairs? We now introduce the strong-coupling gauge formulation that supplies the necessary constraints.

\subsection{Strong Coupling Formulation}\label{sec:intro_strong_coupling}

The comparison above exposes a limitation of conventional BCS mean-field theory: a sign-changing gap on different Fermi surfaces need not satisfy the local exclusion of double occupancy. In a projected $t$--$J$ description, this restriction must remain operative even after the electronic degrees of freedom are represented by spinons and bosons. Slave-boson fractionalization replaces the projection by local constraints, whose Lagrange multipliers appear as emergent gauge fields. The familiar $U(1)$ formulation enforces the local charge constraint, but a pairing mean field must also respect the particle--hole structure of the projected Hilbert space.

The $SU(2)$ formulation of Wen and Lee \cite{WenLee1996}, developed further by Flint \textit{et al.}~\cite{Flintetal2011,FlintColeman2012}, makes this requirement explicit. In a gauge where the boson condensate points along the $b$ direction, $\langle b\rangle=\sqrt{x}$ and $\langle a\rangle=0$, the three local constraints become, at the uniform mean-field level,
\begin{equation}
\label{eq:intro_three_constraints}
\langle n^f_{i\ell}\rangle=1-x,\qquad
\langle f^\dagger_{i\ell\uparrow}f^\dagger_{i\ell\downarrow}\rangle=0,\qquad
\langle f_{i\ell\downarrow}f_{i\ell\uparrow}\rangle=0.
\end{equation}
Here $x$ is the hole density per site, $\ell$ labels the layer, and $f$ denotes a fermionic spinon. The first equation fixes the spinon density; the remaining two are conjugate conditions that eliminate an on-site spinon pair amplitude. Although they represent two equations, together they impose the real and imaginary parts of a single complex pairing constraint. Their associated Lagrange multiplier acts as a self-consistent \textit{Coulomb pseudopotential}, penalizing mean-field pairing states that would otherwise violate the local projection.

This distinction is especially important for the bilayer nickelates. A $d_{x^2-y^2}$ gap has no on-site component by symmetry in the tetragonal limit, whereas a multiband $s_\pm$ state is not automatically protected by its sign change between Fermi sheets. The constraints can therefore alter the competition between intralayer and interlayer pairing. In Section \ref{sec:HundsInducedSuperexchange}, we discuss the relevant solid-state chemistry of La$_3$Ni$_2$O$_7$ and motivate a minimal bilayer $t$-$J$ model. Section~\ref{sec:su2} derives the fractionalized operators, the local $SU(2)$ gauge structure, and the gauge-fixed constraints; Section~\ref{sec:methods} shows how the corresponding Lagrange multipliers enter the bilayer saddle-point equations and pairing instability. This is followed by a discussion of our main results in Section \ref{sec:results}, followed by a discussion in Section \ref{sec:discussions}.

\begin{figure*}
    \centering
    \includegraphics[width=0.45\linewidth]{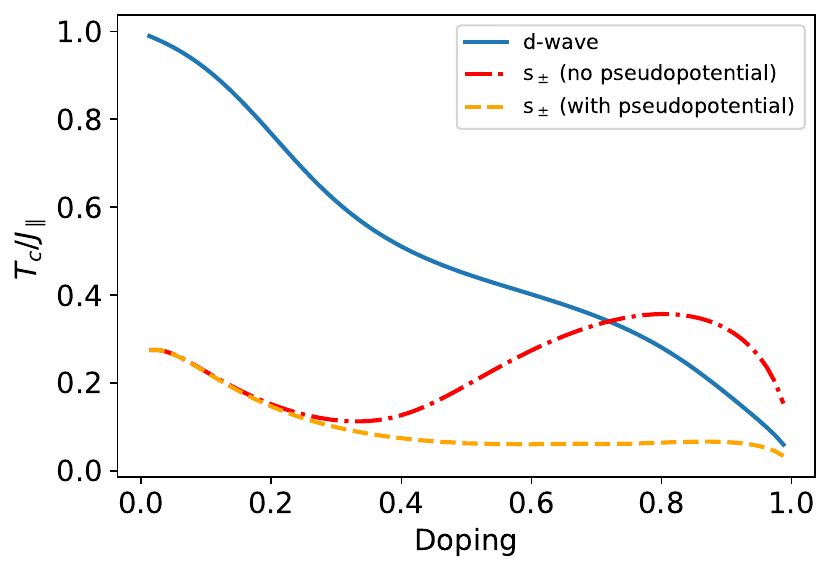}
    \hfill
    \includegraphics[width=0.45\linewidth]{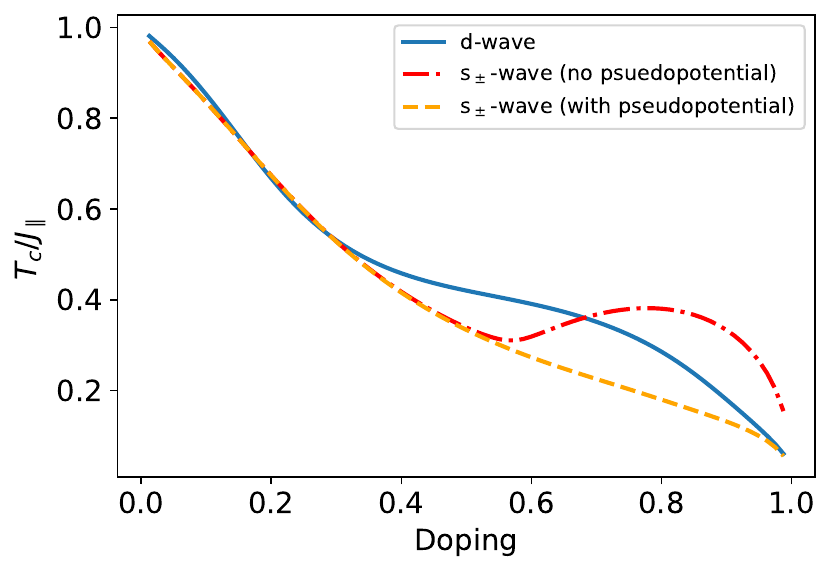}
    \caption{The $T_c$ for $d$-wave vs $s_{\pm}$ wave for two different regimes of coupling strengths. We choose two representative regimes: small $J_\perp$ ($=0.5 J_\parallel$) (left) and large $J_\perp$ (=$J_\parallel$) (right)}
    \label{fig:Tc}
\end{figure*}

\section{Solid State Chemistry of \texorpdfstring{$La_3Ni_2O_7$}{La3Ni2O7} - Induced Superexchange and Minimal Model}\label{sec:HundsInducedSuperexchange}

\begin{figure}
    \centering
    \includegraphics[width=0.95\linewidth]{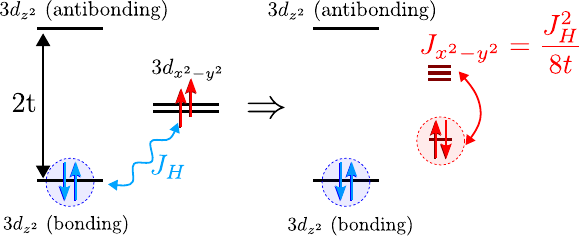}
    \caption{Induced superexchange in the $3d_{x^2 - y^2}$ orbitals. The $3d_{z^2}$ bonding orbital electrons are coupled to the nearly degenerate $3d_{x^2 - y^2}$ orbitals via Hund's coupling (left), which induces the superexchange via virtual fluctuations to the $3d_{z^2}$ antibonding orbital (right)}
    \label{fig:HundsInduced}
\end{figure}

%The strong coupling physics of the cuprates is captured by the $t$-$J$ model, wherein the strong Coulomb repulsion induced an antiferromagnetic superexchange between neighboring Cu sites. 
In the cuprates, in-plane antiferromagnetic superexchange governs the motion of doped Zhang--Rice singlets \cite{ZhangRice1988}. In our model of bilayer La$_3$Ni$_2$O$_7$, the $3d_{x^2-y^2}$ orbitals support both in-plane exchange, generated by virtual hopping between neighboring Ni sites, and interlayer exchange, induced by Hund's coupling to the $3d_{z^2}$ orbitals (Fig.~\ref{fig:HundsInduced}). For the half-filled $3d_{x^2-y^2}$ orbitals, the latter is
\begin{equation}
    J_{\perp} = \frac{J_H^2}{8t_\perp},
\end{equation}
where $H_{\mathrm{Hund}}=-J_H\mathbf S_{x^2-y^2}\cdot\mathbf S_{z^2}$ and $t_\perp$ denotes the interlayer $3d_{z^2}$ hopping (see Appendix \ref{app:hunds_induced} for a detailed derivation). Taking $J_H\simeq0.6$--$1$~eV and a $3d_{z^2}$ bonding--antibonding splitting $2t_\perp\simeq1.2$--$1.6$~eV gives $J^{\rm phys}_\perp\simeq0.05$--$0.2$~eV. For comparison, the in-plane exchange is $J^{\rm phys}_\parallel=4t_\parallel^2/U\simeq0.3$--$0.5$~eV. Thus $J^{\rm phys}_\perp$ is smaller but comparable to $J^{\rm phys}_\parallel$; both estimates depend on pressure and strain \cite{ChristianssonWerner2023,ZhangElbio2023}.

With these considerations in mind, we adopt a projected bilayer $t$--$J$ model.  Let $i$ label an in-plane unit cell, $\ell=1,2$ its two layers, and $X^{ab}_{i\ell}=|a\rangle\langle b|$ the Hubbard operators on the local three-state space $\ket{0},\ket\uparrow,\ket{\downarrow}$.  We use the Pauli-normalized local moment $\bm\sigma_{i\ell}=\sum_{ss'}X^{ss'}_{i\ell}\boldsymbol\sigma_{ss'}=2\mathbf S_{i\ell}$. Throughout the following symplectic decoupling, $J_{\parallel,\perp}$ denote the coefficients multiplying $\bm\sigma_i\cdot\bm\sigma_j$ (the factor conventionally written $1/N$ has been absorbed into $J$). Consequently $J^{\rm phys}_{\parallel,\perp}=4J_{\parallel,\perp}$ when $J^{\rm phys}$ multiplies $\mathbf S_i\cdot\mathbf S_j$. This choice gives the same coupling $J$ in the two symplectic bond channels. The real-space Hamiltonian is
\begin{equation}\label{eq:bilayer-tJ}
\begin{aligned}
H={}&-\sum_{ij,\ell,\sigma}t^{\parallel}_{ij}X^{\sigma0}_{i\ell}X^{0\sigma}_{j\ell}
-\sum_{ij,\sigma}t^{\perp}_{ij}\bigl(X^{\sigma0}_{i1}X^{0\sigma}_{j2}+\mathrm{H.c.}\bigr)\\
&+J_\parallel\sum_{\langle ij\rangle,\ell}\bm\sigma_{i\ell}\cdot\bm\sigma_{j\ell}
+J_\perp\sum_i\bm\sigma_{i1}\cdot\bm\sigma_{i2}.
\end{aligned}
\end{equation}
The hopping matrices are Hermitian, $t^{\parallel}_{ji}=(t^{\parallel}_{ij})^*$; the first sum includes both hopping directions.  The interlayer hopping need not be restricted to vertical rungs: its in-plane displacement dependence is required by the momentum-dependent bilayer splitting used below.  Our convention is $t^{\parallel}_{i,i\pm\hat x}=t^{\parallel}_{i,i\pm\hat y}=t/2$, so that $\epsilon_\parallel(\mathbf k)=-t(\cos k_x+\cos k_y)$.  We choose $t^\perp_{ij}$ such that its Fourier transform obeys $-t^\perp(\mathbf k)=t'(\cos k_x-\cos k_y)^2$.  Explicitly, its nonzero Fourier coefficients are $t^\perp_{\mathbf0}=-t'$, $t^\perp_{\pm2\hat x}=t^\perp_{\pm2\hat y}=-t'/4$, and $t^\perp_{\pm\hat x\pm\hat y}=t'/2$ (each of the four diagonal displacements).  
%This construction includes a vertical hopping and longer-range interlayer transfers; a purely vertical $t_\perp$ would instead produce a constant bilayer splitting.  We retain the exchange interactions on nearest-neighbor in-plane bonds and vertical rungs, as dictated by the superexchange mechanism described above. 
The superexchange estimates above use the conventional physical-spin coefficients $J^{\rm phys}_{\parallel,\perp}$; the $J$ entering the Pauli-normalized Hamiltonian and the Hubbard--Stratonovich action is one quarter of these values. Density-density pieces $-J_{ij}n_i n_j/4$, often included in the canonical $t$--$J$ Hamiltonian, are omitted here.

For comparison with the band structure and with the subsequent mean-field equations, Fourier transforming the projected hopping as a whole, rather than Fourier transforming its constituent slave particles, gives the single-particle Bloch Hamiltonian
\begin{equation}\label{eq:kinetic_ham}
\mathcal H(\mathbf k)=\begin{pmatrix}\epsilon_\parallel(\mathbf k)&\epsilon_\perp(\mathbf k)\\\epsilon_\perp(\mathbf k)&\epsilon_\parallel(\mathbf k)\end{pmatrix},
\end{equation}
where $\epsilon_\parallel=-t(\cos k_x+\cos k_y)$ and $\epsilon_\perp=t'(\cos k_x-\cos k_y)^2$.  The real-space form, Eq.~\eqref{eq:bilayer-tJ}, is the starting point for the fractionalization below.

This replicates the qualitative features of the $\alpha$- and $\beta$-pockets on the Fermi surface as seen in the ARPES data \cite{YangZhou2024, HuYao2025}. In our perspective of the strong coupling model, since only the $3d_{x^2 - y^2}$ orbitals participate in the formation of the RVB state, the system is at a filling of 1/4 (one electron for every two orbitals at each site), and is consequently at a hole doping of 0.5 --- zero doping corresponds to two electrons per site (one electron per orbital) that are localized, and then half a hole is added to each nickel ion. This effective model therefore treats La$_3$Ni$_2$O$_7$ as a heavily hole-doped cuprate system.

\section{Wen--Lee path integral formulation of the model}\label{sec:su2}
We combine the in-plane position and layer label into a single site index $j\equiv(\mathbf R_j,\ell)$, with $\ell=1,2$. A bond $\langle ij\rangle$ may therefore be either an in-plane nearest-neighbor bond or an interlayer rung; its exchange coupling is $J_{ij}=J_\parallel$ or $J_\perp$, respectively. Following Wen and Lee and the symplectic formulation of Flint \textit{et al.}~\cite{WenLee1996,FlintColeman2012}, we represent the projected electron by a fermionic spinon and a two-component charge boson. Let us define
\begin{equation}\label{eq:wl-spinors}
\psi_{j\sigma}=\begin{pmatrix}f_{j\sigma}\\ \tilde\sigma f^\dagger_{j,-\sigma}\end{pmatrix},\ 
B_j=\begin{pmatrix}b_j\\a_j\end{pmatrix},\quad
\tilde\uparrow=+1,\ \tilde\downarrow=-1.
\end{equation}
The physical projected electron is the gauge-singlet composite
\begin{equation}\label{eq:wl-electron}
X_j^{\sigma0}=\psi^\dagger_{j\sigma}B_j
=f^\dagger_{j\sigma}b_j+\tilde\sigma f_{j,-\sigma}a_j,
\quad X_j^{0\sigma}=(X_j^{\sigma0})\dg.
\end{equation}
Note that the two bosons do not describe independent holes, they are required to encode the full SU(2) gauge freedom.
The Pauli-normalized spin density used in Eq.~\eqref{eq:bilayer-tJ} becomes, in the projected spinon representation,
\begin{equation}\label{eq:wl-spin}
\bm{\hat \sigma}_j=\sum_{\alpha, \beta }f^\dagger_{j\alpha}\boldsymbol\sigma_{\alpha \beta }f_{j\beta},\qquad \mathbf S_j=\tfrac12\bm{\hat \sigma}_j.
\end{equation}
The three generators of the local $SU(2)$ redundancy may be expressed as
\begin{subequations}\label{eq:wl-constraints}
\begin{align}
Q_j^3&=n^f_j+b_j^\dagger b_j-a_j^\dagger a_j-1=0,\label{eq:num_constraint}\\
Q_j^+&=f^\dagger_{j\uparrow}f^\dagger_{j\downarrow}+b_j^\dagger a_j=0,\label{eq:pair_constraint_1}\\
Q_j^-&=(Q_j^+)^\dagger=0.\label{eq:pair_constraint_2}
\end{align}
\end{subequations}
The longitudinal constraint fixes the charge, while the transverse constraints remove unphysical on-site spinon pairing. All three are needed to recover the physical projected Hilbert space.

In imaginary time the constraints can be incorporated into the covariant derivatives of the doublets. With $\bm\tau$ the Pauli matrices acting on the gauge (Nambu) indices, define $w_j=\mathbf w_j\cdot\bm\tau$. One can then, in terms of these fractionalized fields, express the path integral as
\begin{widetext}
\begin{equation}\label{eq:wl-lagrangian}
Z=\int\!\mathcal D[\bar\psi,\psi,B^\dagger,B, w]\,e^{-S},\quad
S=\int_0^\beta\!d\tau\,\mathcal L,\qquad
\mathcal L=\frac12\sum_{j\sigma}\psi^\dagger_{j\sigma}(\partial_\tau+w_j)\psi_{j\sigma}
+\sum_j B_j^\dagger(\partial_\tau+w_j)B_j+H_{tJ}.
\end{equation}
\end{widetext}
Here $H_{tJ}$ denotes Eq.~\eqref{eq:bilayer-tJ} expressed through the fractionalized fields. The factor $1/2$ avoids double counting the two Nambu-related spinors; the precise normalization and constant shift of the field are fixed by the constraints in Eqs.~\eqref{eq:wl-constraints}. In particular, $w_j= \vec w_j\cdot \vec \tau$ is a traceless gauge connection that couples to the isospin of the $f$-electrons, imposing the constraints $\psi\dg\vec \tau \psi  + B^\dagger\vec\tau B= 0$. Under a site- and time-dependent $g_j(\tau)\in SU(2)$,
\begin{eqnarray}\label{eq:wl-gauge}
\psi_{j\sigma}&\longrightarrow& g_j\psi_{j\sigma},\quad
B_j\longrightarrow g_j B_j,\cr
w_j&\longrightarrow &g_jw_jg_j^\dagger+g_j\partial_{\tau} g_j^\dagger,
\end{eqnarray}
the composite $\psi^\dagger_{j\sigma}B_j$ and the full action are invariant. In a convention where the Lagrange multipliers are real, the Euclidean temporal connection is understood with the corresponding factor of $i$; we absorb this into $w_j$ here.

%\paragraph{Exchange decoupling.}
To decouple the spins in the exchange term, we use the symplectic completeness relationship of Sp(2N)~\cite{FlintColeman2012}. For even spin-component number $N$, define $\epsilon_{\sigma\sigma'}=\tilde\sigma\delta_{\sigma,-\sigma'}$ and generators $\mathcal S^a$ normalized by $\operatorname{Tr}(\mathcal S^a\mathcal S^b)=2\delta^{ab}$. The symplectic projector $P_S A=(A-\epsilon A^T\epsilon^T)/2$ gives
\begin{equation}\label{eq:spn-completeness}
\sum_a\mathcal S^a_{\alpha\beta}\mathcal S^a_{\gamma\delta}
=\delta_{\alpha\delta}\delta_{\beta\gamma}-\epsilon_{\alpha\gamma}\epsilon_{\beta\delta}.
\end{equation}
At the physical two-component point the generators are Pauli matrices. We do \textit{not} take a one-component spin representation: instead, we absorb the conventional prefactor $1/N$ into the exchange constant $J_{ij}$, so no explicit $N$ appears in the physical model. Inserting Eq.~\eqref{eq:spn-completeness} into $J_{ij}\bm\sigma_i\cdot\bm\sigma_j$ and normal ordering operators on different sites gives the symplectic exchange form
\begin{eqnarray}\label{eq:wl-fierz}
H_J&=& 
 - \hskip -2mm\sum_{\langle ij\rangle, \sigma, \sigma'}\hskip -4mm J_{ij}\biggl[
 f^\dagger_{j\sigma}f_{i\sigma}f^\dagger_{i\sigma'}f_{j\sigma'}
+ \tilde\sigma\tilde\sigma' f^\dagger_{i\sigma}f^\dagger_{j,-\sigma}f_{j,-\sigma'}f_{i\sigma'}\biggr].\cr&&
\end{eqnarray}
If we introduce oriented hopping and pairing  operators
\begin{equation}\label{eq:spn-bond-operators}
D_{ji}=\sum_\sigma f^\dagger_{j\sigma}f_{i\sigma},\qquad
P_{ji}=\sum_\sigma\tilde\sigma f_{j,-\sigma}f_{i\sigma}.
\end{equation}
so that  Eq.~\eqref{eq:wl-fierz} becomes
\begin{equation}\label{eq:spn-bond-hamiltonian}
H_J=-\sum_{\langle ij\rangle}J_{ij}(D_{ji}D_{ji}^\dagger+P_{ji}^\dagger P_{ji}),
\end{equation}
where the displayed operator order follows the normal-ordering convention in Eq.~\eqref{eq:wl-fierz}. The two channels have the same coefficient $J_{ij}$, with $J_{ij}=J_\parallel$ on in-plane bonds and $J_\perp$ on the bonds between the bilayer sites. For each bond we introduce complex Hubbard--Stratonovich fields $V_{ij}$ and $\Delta_{ij}$ conjugate to $D_{ji}$ and $P_{ji}$. Following standard procedure, completing the square in the hopping channel gives
\begin{equation}\label{eq:wl-hs-square}
-J_{ij}D_{ji}D_{ji}^\dagger\ \longrightarrow\
\frac{|V_{ij}|^2}{J_{ij}}-V_{ij}^*D_{ji}-V_{ij}D_{ji}^\dagger,
\end{equation}
and the pairing channel analogously gives $|\Delta_{ij}|^2/J_{ij}-\Delta_{ij}^*P_{ji}-\Delta_{ij}P_{ji}^\dagger$. Varying the auxiliary fields near the saddle point yields $V_{ij}=J_{ij}\langle D_{ji}\rangle$ and $\Delta_{ij}=J_{ij}\langle P_{ji}\rangle$. 
%Thus the Hubbard--Stratonovich quadratic costs follow directly from the same Pauli-normalized $J_{ij}$ as the microscopic exchange; no independent coupling $g_{ij}$ is introduced. The detailed conversion from bond fields to the momentum-space amplitudes must account for bond multiplicity and the normalization of the bilayer Nambu basis.
In Nambu form these channels assemble into the link matrix
\begin{equation}\label{eq:bond-decoupling}
U_{ij}=\begin{pmatrix}V_{ij}&\Delta_{ij}\\\Delta^*_{ij}&-V^*_{ij}\end{pmatrix},\qquad
U_{ij}\longrightarrow g_iU_{ij}g_j^\dagger.
\end{equation}
The link transformation ensures that the quadratic spinon term is gauge covariant.

For a doped homogeneous saddle point, the boson is assumed to condense, $\langle B_j\rangle\ne0$. A local $SU(2)$ transformation rotates this doublet into the $b$ direction; we choose its remaining phase to be real:
\begin{equation}\label{eq:gutz-gauge}
\langle B_j\rangle=\begin{pmatrix}\sqrt{x}\\0\end{pmatrix},\qquad
\langle a_j\rangle=0,\quad\langle b_j\rangle=\sqrt{x}.
\end{equation}
With $j=(\mathbf R_j,\ell)$, the projected hopping becomes
\begin{eqnarray}\label{eq:wl-bilayer-hopping}
H_{t,\mathrm{MF}}&=&-x\sum_{\mathbf R,\mathbf R',\ell,\sigma}t^\parallel_{\mathbf R\mathbf R'}f^\dagger_{\mathbf R\ell\sigma}f_{\mathbf R'\ell\sigma}\cr
-&&x\sum_{\mathbf R,\mathbf R',\sigma}t^\perp_{\mathbf R\mathbf R'}\bigl(f^\dagger_{\mathbf R1\sigma}f_{\mathbf R'2\sigma}+\mathrm{H.c.}\bigr).
\end{eqnarray}
The remaining mean-field constraints are, with the combined site label restored,
\begin{subequations}\label{eq:wl-mf-constraints}
\begin{align}
\langle n^f_j\rangle&=1-x,\label{eq:num_constraint_mf}\\
\langle f^\dagger_{j\uparrow}f^\dagger_{j\downarrow}\rangle&=0,\label{eq:pair_constraint_1_mf}\\
\langle f_{j\downarrow}f_{j\uparrow}\rangle&=0.\label{eq:pair_constraint_2_mf}
\end{align}
\end{subequations}
For the real, time-reversal-symmetric ansatz, the transverse components of the temporal field reduce to an on-site pair multiplier $w$, while its longitudinal component gives $\lambda$. The gauge-fixed constraint Hamiltonian is
\begin{equation}\label{eq:lagrange_1}
H_C=\sum_j\!\left\{\lambda[n_j^f-(1-x)]+w[f^\dagger_{j\uparrow}f^\dagger_{j\downarrow}+\mathrm{H.c.}]\right\}.
\end{equation}
Combining the hopping, exchange links and constraints, the full real-space saddle-point Lagrangian can be summarized as
\begin{widetext}
\begin{equation}\label{eq:wl-full-mf-lagrangian}
\boxed{\begin{aligned}
\mathcal L_{\mathrm{MF}}={}&\sum_{j\sigma}f^\dagger_{j\sigma}\partial_\tau f_{j\sigma}
-x\sum_{\mathbf R\mathbf R',\ell,\sigma}t^\parallel_{\mathbf R\mathbf R'}f^\dagger_{\mathbf R\ell\sigma}f_{\mathbf R'\ell\sigma}
-x\sum_{\mathbf R\mathbf R',\sigma}t^\perp_{\mathbf R\mathbf R'}(f^\dagger_{\mathbf R1\sigma}f_{\mathbf R'2\sigma}+\mathrm{H.c.})\\
&-\sum_{\langle ij\rangle}\bigl[V_{ij}^*D_{ij}+V_{ij}D^\dagger_{ij}+\Delta_{ij}^*P_{ij}+\Delta_{ij}P^\dagger_{ij}\bigr]
+\sum_{\langle ij\rangle}\frac{|V_{ij}|^2+|\Delta_{ij}|^2}{J_{ij}}
+\sum_j\bigl\{\lambda[n^f_j-(1-x)]+[ w f^\dagger_{j\uparrow}f^\dagger_{j\downarrow}+\mathrm{H.c.}]\bigr\}.
\end{aligned}}
\end{equation}
\end{widetext}
The sums over $\langle ij\rangle$ include the in-plane bonds and interlayer rungs, with their respective couplings. We use equation~\eqref{eq:wl-full-mf-lagrangian} as the real-space starting point for Section~\ref{sec:methods}, where we transform the link fields to momentum space and derive the bilayer Bogoliubov--de Gennes spectrum and saddle-point equations. While the longitudinal Lagrange multiplier $\lambda$ constrains the spinon filling to the doping $x$, The on-site field $w$ enforces the vanishing of the net anomalous spinon amplitude and thus acts as a strong-coupling \textit{Coulomb pair pseudopotential}.

\section{Methods: bilayer saddle point and pairing instability}\label{sec:methods}
We now specialize the real-space saddle-point Lagrangian, Eq.~\eqref{eq:wl-full-mf-lagrangian}, to homogeneous intralayer and interlayer link fields. The projected hopping and the Hubbard--Stratonovich decoupling have already been derived in Section~\ref{sec:su2}; here we fix the momentum-space conventions and obtain the bilayer quasiparticle spectrum. The uniform constraints are those in Eq.~\eqref{eq:lagrange_1}.

Let us take $V_x=V_y=V_\parallel/2$ in each layer, $V_z=V_\perp$ on each rung, and real in-plane pair fields $\Delta_x,\Delta_y$.  Their Fourier transform is
\begin{equation}\label{eq:formfactor-normalization}
2\Delta_x\cos k_x+2\Delta_y\cos k_y
=\Delta_s\gamma_s(\mathbf k)+\Delta_d\gamma_d(\mathbf k),
\end{equation}
where $\Delta_s=\Delta_x+\Delta_y$, $\Delta_d=\Delta_x-\Delta_y$, $\gamma_s=\cos k_x+\cos k_y$ and $\gamma_d=\cos k_x-\cos k_y$.  If $\mathcal N_s$ denotes the number of \textit{bilayer unit cells}, the two layers together give the bond-field cost $2\mathcal N_s(\Delta_x^2+\Delta_y^2)/J_\parallel=\mathcal N_s(\Delta_s^2+\Delta_d^2)/J_\parallel$ in this normalization.  The exchange constants are Pauli-normalized, as in Section~\ref{sec:su2}; bond counting fixes their relation to the momentum-space amplitudes. We compare separate symmetry sectors, $a=s$ or $a=d$, rather than assuming a state that spontaneously mixes them.  The $s$ sector allows both $\Delta_s$ and a rung singlet $\Delta_z$, producing bonding/antibonding gaps $\Delta_s\gamma_s\pm\Delta_z$; the $d$ sector has $\Delta_d\gamma_d$, which is orthogonal to the on-site $s$-wave constraint by lattice symmetry.

To take care of the Nambu and interlayer degrees of freedom and to allow ourselves to write the final mean-field Bogoliubov-de Gennes (BdG) Hamiltonian with cleaner notation, we introduce the four-component spinor 
\begin{equation}
    \tilde f_{\bm k} = 
    \begin{pmatrix}
        f_{\bm k, 1,\uparrow} \\
        f_{\bm k, 2,\uparrow} \\
        f^\dagger_{-\bm k, 1,\downarrow} \\
        f^\dagger_{-\bm k, 2,\downarrow} \\
    \end{pmatrix}
\end{equation}
spanned by the Pauli operators $\bm \alpha$ and $\bm \tau$. $\bm{\alpha}$ acts on the layer degree of freedom on the four-component space, where $\alpha_0$ is the identity ($\mathbb{1}_{4\times 4}$) and 
\begin{equation}
    \alpha_1 = 
    \begin{pmatrix}
        \sigma_1 & 0\\
        0 & \sigma_1
    \end{pmatrix}.
\end{equation}
Similarly, the Nambu isospin degrees of freedom are acted on by $\bm \tau$,

\begin{equation}
    \tau_1 = 
    \begin{pmatrix}
        0 & \mathbb{1} \\
        \mathbb{1} & 0
    \end{pmatrix}
    ,\quad
    \tau_3 = 
    \begin{pmatrix}
        \mathbb{1} & 0\\
        0 & -\mathbb{1}
    \end{pmatrix}.
\end{equation}

With these operators defined, we can rewrite the mean-field Hamiltonian as shown below:

\begin{widetext}
\begin{subequations}\label{eq:meanfield}
    \begin{align}
        H_{\mathrm{MF}} &= H_{f,\mathrm{MF}} + H_{c,\mathrm{MF}}, \nonumber\\
        H_{f,\mathrm{MF}} &= \sum_{{\bm k}}\tilde{f}^\dagger_{\bm k} 
        \left[
        \left(x\mathcal{H}(\bm k) + V_{||}(c_x + c_y) \alpha_0 + V_\perp \alpha_1 \right)\tau_3 + \left(\sum_a \Delta_a \gamma_a(\bm k)\alpha_0 + \Delta_z\alpha_1 \right)\tau_1\right]
        \tilde{f}_{\bm k} \nonumber\\
        &\hspace{16em}+\mathcal{N}_s \left(\frac{\sum_a|\Delta_{a}|^2 + |V_{||}|^2}{J_\parallel} + \frac{|\Delta_z|^2 + |V_z|^2}{J_\perp}\right), \\
        H_{c,\mathrm{MF}} &= \sum_{{\bm k}} \tilde f^\dagger_{\bm k}\left(\lambda\alpha_0\tau_3 + w\alpha_0\tau_1\right)
        \tilde f_{\bm k} + 2\mathcal{N}_s\lambda x
    \end{align}
\end{subequations}
\end{widetext}
Here $\mathcal N_s$ is the number of {bilayer unit cells}, so that there are $2\mathcal N_s$ physical nickel sites. The constraint term $2\mathcal N_s\lambda x$ is required by the chosen Nambu convention. 

%Indeed, with $\alpha_\pm=E_\parallel\pm E_\perp$ and $E_\pm=(\alpha_\pm^2+\beta_\pm^2)^{1/2}$, differentiation of the quasiparticle contribution gives $\partial F_{\rm qp}/\partial\lambda=-\sum_{\mathbf k,\pm}\tanh(E_\pm/2T)\alpha_\pm/E_\pm$. Consequently $\partial F/\partial\lambda=0$ gives $\int_{\mathbf k}\sum_\pm\tanh(E_\pm/2T)\alpha_\pm/E_\pm=2x$, as required. The factor of two counts the two layers.\par

The free energy for the above Hamiltonian is calculated by finding the eigenvalues of the BdG Hamiltonian, and can be written as

\begin{equation}
    \begin{aligned}
    F &= -2T\sum_{\bm{k},\pm} \ln\left[2 \cosh\left(\frac{E_{\pm}(\bm k)}{2T}\right)\right]\\
    &+ \mathcal{N}_s \left(\frac{\sum_a|\Delta_{a}|^2 + |V_{||}|^2}{J_\parallel} + \frac{|\Delta_z|^2 + |V_z|^2}{J_\perp}\right)\\
    &+2\mathcal{N}_s\lambda x,
    \end{aligned}
\end{equation}
where the two Bogoliubov quasiparticle bands (bonding and antibonding) have the energies
\begin{eqnarray}
    E_{\pm}(\bm k) &=& \big[(\alpha_\pm (\bm k))^2 + (\beta_\pm (\bm k))^2 \big]^{0.5}.
    \cr
    \alpha_\pm (\bm k)&=& E_\parallel(\bm k) \pm E_{\perp}\cr
    \beta_\pm (\bm k)&=& w +\sum_a \Delta_a \gamma_a(\bm k) \pm \Delta_z
\end{eqnarray}
The energies $E_\parallel,\ E_{\perp}$ are a combination of the true dispersion on the bands combined with the dispersion generated from decoupling the superexchange into the particle-hole channel.

\begin{align}
    E_\parallel(\bm k) &= x\epsilon_\parallel(\bm k) + V_\parallel (c_x + c_y) + \lambda, \\
    E_\perp(\bm k) &= x\epsilon_\perp(\bm k)+ V_{\perp}.
\end{align}
The solution to this mean-field free energy lies at the saddle point. By extremizing the free energy with respect to $\lambda$, $V_\parallel$, $V_\perp$, $w$, $\Delta_a$ and $\Delta_z$ respectively, we obtain the following self-consistency solutions.

\begin{alignat}{2}
    &\int_{\bm k} \sum_\pm 
    \frac{\tanh\left({E_\pm(\bm k)}/{2T}\right)}{E_\pm(k)}\alpha_\pm (\bm k) &= 2x\\
    &\int_{\bm k} \sum_\pm \frac{\tanh\left({E_\pm(\bm k)}/{2T}\right)}{E_\pm(k)}\alpha_\pm (\bm k)(c_x+c_y) &= \frac{2V_\parallel}{J_\parallel}\\
    &\int_{\bm k} \sum_\pm (\pm) \frac{\tanh\left({E_\pm(\bm k)}/{2T}\right)}{E_\pm(k)}\alpha_\pm (\bm k)&= \frac{2V_\perp}{J_\perp} \\
    &\int_{\bm k} \sum_\pm \frac{\tanh\left({E_\pm(\bm k)}/{2T}\right)}{E_\pm(k)}\beta_\pm (\bm k) &= 0 \\
    &\int_{\bm k} \sum_\pm \frac{\tanh\left({E_\pm(\bm k)}/{2T}\right)}{E_\pm(k)}\beta_\pm (\bm k)\gamma_a(\bm k) &=  \frac{\Delta_a}{J_\parallel} \\
    &\int_{\bm k} \sum_\pm (\pm) \frac{\tanh\left({E_\pm(\bm k)}/{2T}\right)}{E_\pm(k)}\beta_\pm (\bm k) &=  \frac{\Delta_z}{J_\perp}
\end{alignat}
where the integral $\int_{\bm k} = (2\pi)^{-2}\int_{BZ} (d^2k) = \frac{1}{\mathcal{N}_s}\sum_{\bm k}$ is the average over the Brillouin zone (discrete or continuous). At the critical temperature, the pairing amplitudes and the pair repulsion constraint vanish, so $E_\pm(\bm k)=|\alpha_\pm(\bm k)|$ (not $\alpha_\pm$ without an absolute value). In expressions containing $\alpha_\pm/E_\pm$, the combination is $\operatorname{sgn}(\alpha_\pm)$, so that $\tanh(E_\pm/2T)\alpha_\pm/E_\pm=\tanh(\alpha_\pm/2T)$. The first three equations are then simplified to

\begin{alignat}{3}
    &\int_{\bm k} \sum_\pm 
    \tanh\left(\frac{\alpha_\pm(\bm k)}{2T_c}\right)&= {2x}\\
    &\int_{\bm k} \sum_\pm {\tanh\left(\frac{\alpha_\pm(\bm k)}{2T_c}\right)}(c_x+c_y) &= \frac{2V_\parallel}{J_\parallel}\\
    &\int_{\bm k} \sum_\pm (\pm) {\tanh\left(\frac{\alpha_\pm(\bm k)}{2T_c}\right)}&= \frac{2V_\perp}{J_\perp}
\end{alignat}
where $\beta_c$ is the critical inverse temperature, and the second three equations are replaced with the condition $\det\bm\chi = 0$, where $\bm \chi$ is the susceptibility matrix.

\begin{equation} \label{eq:susmatrix}
    \bm \chi = 
    \begin{pmatrix}
        \chi_{ww} & \chi_{wa} & \chi_{wz} \\
        \chi_{wa} & \chi_{aa} - \frac{1}{J_\parallel} & \chi_{az} \\
        \chi_{wz} & \chi_{az} & \chi_{zz} - \frac{1}{J_\perp}\\
    \end{pmatrix}
\end{equation}

The elements of the susceptibility matrix are:

\begin{align}
    \chi_{ww} = &\int_{\bm k} \sum_\pm \frac{\tanh\left({\alpha_\pm(\bm k)}/{2T_c}\right)}{\alpha_\pm(k)} =\chi_{zz}\\
    \chi_{wa} = &\int_{\bm k} \sum_\pm \frac{\tanh\left({\alpha_\pm(\bm k)}/{2T_c}\right)}{\alpha_\pm(k)}\gamma_a(\bm k) \\
    \chi_{wz} = &\int_{\bm k} \sum_\pm (\pm)\frac{\tanh\left({\alpha_\pm(\bm k)}/{2T_c}\right)}{\alpha_\pm(k)} \\
    \chi_{aa} = &\int_{\bm k} \sum_\pm \frac{\tanh\left({\alpha_\pm(\bm k)}/{2T_c}\right)}{\alpha_\pm(k)} (\gamma_a(\bm k))^2\\
    \chi_{az} = &\int_{\bm k} \sum_\pm (\pm)\frac{\tanh\left({\alpha_\pm(\bm k)}/{2T_c}\right)}{\alpha_\pm(k)} \gamma_a(\bm k)
\end{align}

These self-consistency conditions are solved simultaneously to find the critical temperature $T_c$. At the superconducting transition, the determinant of the susceptibility $\det{\bm\chi}$ changes sign, corresponding to a sign change in one of the eigenvalues of $\chi$. By finding the eigenvector corresponding to this unstable eigenvalue, we can obtain the relative proportions of extended-s ($\Delta_s(c_x + c_y)$) to interlayer ($\Delta_z$) pairing, allowing us to plot the gap at the spinon Fermi surface.

\begin{figure}[ht]
    \centering
    \includegraphics[width=0.49\linewidth]{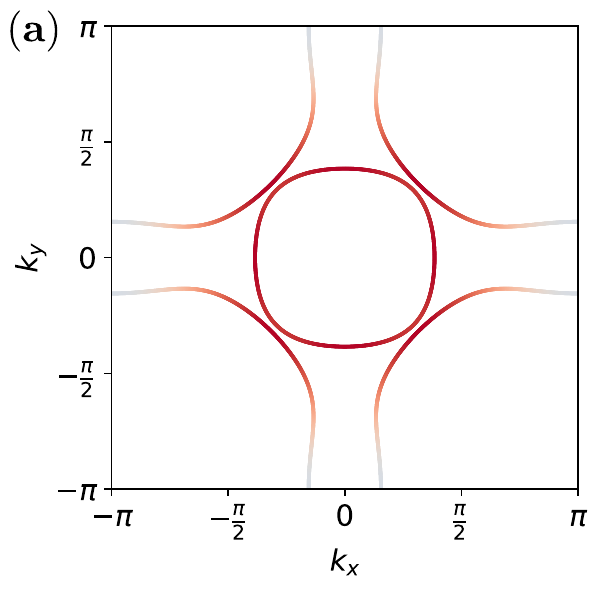}
    \hfill
    \includegraphics[width=0.49\linewidth]{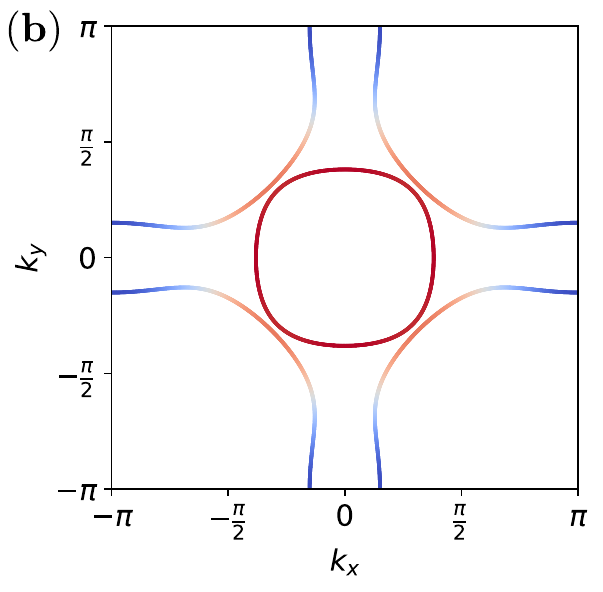}
    \caption{Fermi-surface projection of $s_\pm$ gap as $T\rightarrow T_c$ for $J_\perp/J_\parallel = 1/2$, for intermediate doping. \textbf{(a)}: no Coulomb psuedopotential. \textbf{(b)}: with Coulomb psuedopotential.}
    \label{fig:spm_gap_FS}
\end{figure}

\section{Results} \label{sec:results}

By using the Wen-Lee fractionalization scheme and numerically minimizing the free energy by solving the coupled self-consistency equations, we have found the critical temperature $T_c$ for different values of doping. We present a full sweep for the hole doped regime for completeness; however, as stated before, the bilayer system La$_3$Ni$_2$O$_7$ corresponds to a doping of $x = 0.5$. Our results are summarized in Fig. \ref{fig:Tc}. In both cases, we have chosen values of $J_\perp$ and $J_\parallel$ much smaller than the bandwidth $W$ (in our case, $J_\parallel/W\sim1/8$), but the qualitative takeaways from the results are by and large insensitive to this fact. In the weak $J_\perp$ regime, we chose the value of $J_\perp = \frac{1}{2}J_\parallel$. In this regime, we see that for small doping, the $T_c$ for $d$-wave order far exceeds $s_\pm$ order. In the low doping regime, there is no effect of the Coulomb pseudopotential. This is to be expected, since at half-filling the Coulomb constraint is satisfied automatically. Upon hole doping, the $T_c$ for both $d$-wave and $s_\pm$ fall, but at a doping of $0.25$ a sharp upturn is noticed in the $s_\pm$ $T_c$. This sharp upturn coincides with the pairing gap acquiring the \textit{same sign} on both Fermi surfaces (Fig. \ref{fig:spm_gap_FS}a), leading to a mean-field solution with no nodes, a higher free energy and therefore a higher $T_c$. At even larger doping ($x\approx 0.75$), $s_\pm$ manages to surpass $d$-wave in $T_c$. However, once the Coulomb pseudopotential constraint is added to the solution, the $s_\pm$ $T_c$ is suppressed significantly, never crossing the $d$-wave $T_c$. By imposing the Coulomb constraint, the pair function is forced to change signs between Fermi surfaces (Fig. \ref{fig:spm_gap_FS}b)

We had argued previously in Sec. \ref{sec:HundsInducedSuperexchange} that the induced interlayer superexchange between the $3d_{x^2- y^2}$ orbitals is expected to be smaller than the in-plane superexchange. However, we consider for the sake of completeness the case where $J_\perp$ is of comparable magnitude to $J_\parallel$ --- namely, $J_\perp = J_\parallel$ (Fig. \ref{fig:Tc}, right). In this case, the competition between the two orders is stronger. For low doping, the $T_c$ for $d$-wave is nearly identical to $s_\pm$. This is not that surprising since the interlayer and intralayer couplings are the same strength. This coincidence continues down to moderate doping ($x\sim0.3$), where $d$-wave wins out over $s_\pm$. Note that in both this and the previous plot, the $T_c$ for $d$-wave follows identical trajectories. This is because band renormalization effects on the $d$-wave pairing state are minimal for weaker interlayer coupling. At half-doping, same as before, the $s_\pm$ $T_c$ starts to rise, emerging as the dominant state for $x\gtrsim0.7$. Upon the addition of Coulomb repulsion, the $T_c$ for $s_\pm$ is suppressed below $d$-wave for all doping beyond $x\sim0.7$.

The addition of the Coulomb pseudopotential at the mean-field level, by suppressing onsite pairing, always reduces the $T_c$ of the $s_\pm$ pairing order. One way to think about this that is particularly elucidating is to look at the susceptibility matrix $\bm\chi$ defined in \eqref{eq:susmatrix}. We can rewrite this as

\begin{equation} \label{eq:susmatrix_modified}
    \bm \chi = 
    \begin{pNiceArray}{c|cc}
        \chi_{ww} & \chi_{wa} & \chi_{wz} \\
        \hline
        \chi_{aw} & \Block{2-2}{\bm{\chi_0}} \\
        \chi_{zw} & &\\
    \end{pNiceArray},
\end{equation}
where $\bm{\chi_0}$ is the bare susceptibility without the Coulomb psuedopotential constraint. At $T=T_c$, $\det \bm\chi$ vanishes, and using block matrix properties, one can rewrite this vanishing condition as $\chi_{ww}\det(\bm{\tilde\chi_0})=0$, where

\begin{equation} \label{eq:reduced_suscep}
    \bm{\tilde{\chi}_0} = \bm{\chi_0} - \frac{1}{\chi_{ww}}
    \begin{pmatrix}
        \chi_{wa} \\ \chi_{wz}
    \end{pmatrix}
    \begin{pmatrix}
        \chi_{aw} & \chi_{zw}
    \end{pmatrix}
\end{equation}

We observe that, according to \eqref{eq:reduced_suscep}, the addition of the Coulomb psuedopotential creates an effective susceptibility $\bm{\tilde \chi_0}$ that is the original susceptibility with a positive semidefinite correction. This correction reduces the eigenvalues of the susceptibility matrix, thereby reducing $T_c$ for $s_\pm$ pairing. For $d$-wave pairing, the correction to the bare susceptibility is 0, since the $d$-wave symmetry is completely orthogonal to the Coulomb pseudopotential as well as the dimer pair order $\Delta_z$.

\section{Discussion and Outlook}
\label{sec:discussions}

In this work, we have investigated the influence of strong
local Coulomb repulsion on the competition between $d$-wave
and $s_\pm$ superconductivity in bilayer nickelates.
Using the SU(2) fractionalization of the bilayer $t$-$J$
model, we have shown that the suppression of on-site
pairing introduces a constraint on the superconducting
order parameter that is absent in conventional mean-field
treatments. This constraint suppresses the $s_\pm$
pairing susceptibility through a positive-semidefinite
correction, while leaving the $d$-wave channel unaffected
by symmetry.

The Coulomb constraint does not, however, prohibit
$s_\pm$ superconductivity. Strong interlayer exchange
$J_\perp$ enhances interlayer pairing and can partially
overcome the Coulomb suppression. The resulting competition
is controlled by the balance between intralayer and
interlayer superexchange, subject to the requirement
that the condensate avoid local double occupancy.
For the parameters studied here, the constraint favors
$d$-wave superconductivity over much of the phase
diagram, including the nominal doping $x=0.5$
of La$_3$Ni$_2$O$_7$.

An important motivation for studying the nickelates is
the possibility of identifying the essential ingredients
of high-temperature superconductivity in both the cuprates and nickelates.
Both families share closely related square-planar
transition-metal--oxygen structures, strongly correlated
$d_{x^2-y^2}$ orbitals, and charge fluctuations involving
$d^8$ and $d^9$ configurations. Both also exhibit
strange-metal behavior. Given the rarity of
high-temperature superconductivity, these similarities
provide a compelling motivation to seek a common
microscopic description rather than separate,
material-specific explanations.

Several directions for future work emerge from this
perspective.
\begin{itemize}
\item {\sl Effective low-energy models.}
In the cuprates, Zhang--Rice singlet formation provides
a route to an effective single-band $t$-$J$ model.
In the bilayer nickelates, strong interlayer hybridization
of the $d_{z^2}$ orbitals may play an analogous role
by splitting these states away from the Fermi energy.
Establishing whether this leaves a predominantly
$d_{x^2-y^2}$ low-energy electronic structure is
an important experimental and theoretical challenge.
Comparison with infinite-layer nickelates, where
the orbital hierarchy arises without strong bilayer
hybridization, may help determine whether these
materials admit a common low-energy description.

\item {\sl Experimental tests of pairing symmetry.}
Recent ARPES measurements on ambient-pressure
(La,Pr,Sm)La$_3$Ni$_2$O$_7$ films
\cite{ShenChen2026} reveal nodeless gaps on the
$\beta$ and $\gamma$ Fermi-surface pockets,
challenging the $d$-wave state favored by our
calculations. The additional $\gamma$ pocket,
neglected in our minimal model, may modify the
competition between pairing channels by allowing
cancellations of the local anomalous amplitude
between different bands. Phase-sensitive measurements,
together with comparisons between ambient-pressure
films and pressurized bulk materials, could help
establish whether the observed superconductivity
is consistent with an $s_\pm$ state satisfying
the Coulomb constraint.

\item {\sl Beyond mean-field theory.}
Multiorbital calculations incorporating the
$d_{z^2}$ states and Hund's coupling are needed
to establish the robustness of our results.
Cluster DMFT and related approaches could also
examine how the Coulomb constraint is realized
when the anomalous self-energy acquires momentum,
orbital, and frequency dependence. The central
question is whether these additional degrees of
freedom qualitatively alter the competition between
$d$-wave and $s_\pm$ pairing, or can be integrated
out to recover a simpler effective theory.
\end{itemize}

We conclude by noting that the  discovery of ambient-pressure nickelate superconductivity
opens an exceptional opportunity to compare the cuprates and
nickelates through common experimental and theoretical probes.
It transforms the cuprate problem from the study of a unique
family of materials into a comparative search for universal
principles. The challenge is not to construct separate explanations for each superconductor, but to identify the minimal physics underlying strange-metal behavior and high-temperature superconductivity in a unified family of materials—the cupra-nickelates.

\section{Acknowledgements}

The authors would like to thank Liam L. H. Lau, Indra Gankhuyag, Po-Jui Chen and Andreas Gleis for helpful discussions.  
% Kaustubh - putting this here will, I believe help the renewal.
This work was supported by the Office of Basic Energy Sciences, Material
Sciences and Engineering Division, U.S. Department of Energy (DOE)
under Contract DE-FG02-99ER45790.

\appendix
\begin{comment}
    \begin{figure*}[t]
    \centering
    \includegraphics[width=0.8\linewidth]{figures/HundsInduced_AlternateFig.pdf}
    \caption{\KR{alternate figure for the Hund's induced superexchange showing the intermediate states}}
    \label{fig:HundsInducedAlt}
\end{figure*}
\end{comment}

\section{Calculation for the Hund's induced superexchange in the \texorpdfstring{$3d_{x^2 - y^2}$}{3d x squared - y squared} orbitals of \texorpdfstring{La$_3$Ni$_2$O$_7$}{La3Ni2O7}}\label{app:hunds_induced}

Consider a two-site model (site 1 and 2) with two orbitals on each site labelled by $a$ and $b$ (in our case, the $3d_{z^2}$ and $3d_{x^2 - y^2}$ orbitals in Ni). Assume that the $a$ orbitals have a strong overlap, so the dominant energy scale in the problem is the hopping amplitude $t$.

\begin{equation}
    H_{aa} = -t\sum_\sigma c^\dagger_{1a\sigma}c_{2a\sigma} + c^\dagger_{2a\sigma}c_{1a\sigma}.
\end{equation}

Now assume there is a weaker Hund's coupling between the $a$- and $b$- orbitals on each site.

\begin{equation}
    H_{H} = -J_H (\S_{1a}\cdot \S_{1b} + \S_{2a}\cdot\S_{2b})
\end{equation}

We want to find the induced superexchange between the $b$-orbitals from virtual spin-flip fluctuations from the Hund's exchange. We first recast $H_{aa}$ in terms of bonding and antibonding orbitals.

\begin{align}
    B^\dagger &= \frac{1}{\sqrt{2}}\left(c^\dagger_{1a} + c^\dagger_{2a}\right) \\
    A^\dagger &= \frac{1}{\sqrt{2}}\left(c^\dagger_{1a} - c^\dagger_{2a}\right)
\end{align}

which gives us

\begin{equation}
    \begin{aligned}
    H_{aa} &= -t\sum_\sigma \frac{1}{2}(B^\dagger_\sigma + A^\dagger_\sigma)(B_\sigma - A_\sigma) +\frac{1}{2}(B^\dagger_\sigma - A^\dagger_\sigma)(B_\sigma + A_\sigma) \\
    &= -t\sum_\sigma B^\dagger_\sigma B_\sigma - A^\dagger_\sigma A_\sigma
    \end{aligned}    
\end{equation}
The ground state of the system is the bonding-orbital singlet $\ket{S_B} \equiv B^\dagger_\uparrow B^\dagger_\downarrow \ket{0}$. An excitation to the antibonding orbital costs energy $2t$. Now we rewrite the onsite spin operators for the $a$- orbital in terms of the bonding and antibonding orbitals.

\begin{equation}
\begin{aligned}
    \S_{1a} &= c^\dagger_{1a\sigma}\left(\frac{\bm\sigma}{2}\right)_{\sigma\sigma'}c_{1a\sigma'} \\
    & = \frac{1}{2}(B^\dagger_\sigma + A^\dagger_\sigma)\left(\frac{\bm\sigma}{2}\right)_{\sigma\sigma'}(B_{\sigma'} + A_{\sigma'})\\
    & = \frac{1}{2}(\S_B + \S_A + \underbrace{\left(\frac{\bm\sigma}{2}\right)_{\sigma\sigma'}\left(B^\dagger_\sigma A_{\sigma'} + A^\dagger_\sigma B_{\sigma'}\right)}_{\bm{T}})
\end{aligned}
\end{equation}
Similarly, the other spin operator is defined as

\begin{equation}
    \S_{2a} = \frac{1}{2}(\S_B + \S_A - \bm{T}).
\end{equation}
Here, $\bm{T}$ is the spin-flip operator. The action of the Hund's coupling on $\ket{S_B}$ is clear - both $\S_B$ and $\S_A$ evaluate to zero, so the only remaining contribution is:

\begin{equation}
    H_H = -\frac{J_H}{2}(\bm{T}\cdot(\S_{1b} - \S_{2b}))
\end{equation}
The action of the spin-flip operator on the bonding orbital singlet can be calculated component by component. We can ignore the first terms in $\bm T$ since they include a B creation operator, which acts trivially on the ground state singlet.

\textbf{z-component:}
\begin{equation}
\begin{aligned}
    T_z\ket{S_B} &= \frac{1}{2}(A^\dagger_\uparrow B_\uparrow - A^\dagger_\downarrow B_\downarrow)(B^\dagger_\uparrow B^\dagger_\downarrow)\ket{0} \\
    &= \frac{1}{2}(A^\dagger_\uparrow B^\dagger_\downarrow+A^\dagger_\downarrow B^\dagger_\uparrow)\ket{0} \\
\end{aligned}
\end{equation}

\textbf{x-component:}
\begin{equation}
\begin{aligned}
    T_x\ket{S_B} &= \frac{1}{2}(A^\dagger_\downarrow B_\uparrow + A^\dagger_\uparrow B_\downarrow)(B^\dagger_\uparrow B^\dagger_\downarrow)\ket{0} \\
    &= \frac{1}{2}(A^\dagger_\downarrow B^\dagger_\downarrow -A^\dagger_\uparrow B^\dagger_\uparrow )\ket{0} \\
\end{aligned}
\end{equation}

\textbf{y-component:}
\begin{equation}
\begin{aligned}
    T_y\ket{S_B} &= \frac{1}{2}(iA^\dagger_\downarrow B_\uparrow -i A^\dagger_\uparrow B_\downarrow)(B^\dagger_\uparrow B^\dagger_\downarrow)\ket{0} \\
    &= \frac{1}{2}(iA^\dagger_\downarrow B^\dagger_\downarrow + iA^\dagger_\uparrow B^\dagger_\uparrow )\ket{0} \\
\end{aligned}
\end{equation}

Each of these three states are orthogonal triplets with the same energy above the ground state $(2t)$. The effective coupling between the b-orbitals is

\begin{equation}
    H_{\text{eff}} = \bra{S_B}H_H \frac{1}{E_{S_B} - H_{aa}}H_H\ket{S_B}.
\end{equation}

The cross terms vanish ($\langle T_xT_y\rangle = \langle T_zT_y\rangle = \langle T_xT_z\rangle = 0$), and the only contributions left are

\begin{equation}
    \bra{S_B}T_x^2\ket{S_B} = \bra{S_B}T_y^2\ket{S_B}= \bra{S_B}T_z^2\ket{S_B} = \frac{1}{2}
\end{equation}

This leaves us with

\begin{equation}
\begin{split}
    H_{\text{eff}} &= \frac{1}{2}\left(-\frac{J_H}{2}\right)^2 \left(\frac{1}{0-2t}\right)(\S_{1b} - \S_{2b})^2\\
    & = -\frac{1}{16}\frac{J_H^2}{t}(\S_{1b} - \S_{2b})^2\\
    & \equiv \frac{1}{8}\frac{J_H^2}{t}\S_{1b}\cdot\S_{2b}
\end{split}
\end{equation}
We have dropped irrelevant constant terms in the last line. This gives us an effective effective AFM superexchange between the $b$-orbitals of $J_b = \frac{J_H^2}{8t}$.

\begin{comment}
    \section{Self-consistency equations and \texorpdfstring{$T_c$}{Tc} calculation details for the bilayer \texorpdfstring{$t$-$J$}{t-J} model}\label{app:self_consistency_tc}

The two Bogoliubov quasiparticle bands have the energies

\begin{equation}
    E_{\pm}(\bm k) = \sqrt{(\underbrace{E_\parallel \pm E_{\perp}}_{\alpha_\pm (\bm k)})^2 + (\underbrace{w + \Delta_a \gamma_a(\bm k) \pm \Delta_z}_{\beta_\pm (\bm k)})^2}.
\end{equation}
The energies $E_\parallel,\ E_{\perp}$ are a combination of the true dispersion on the bands combined with the dispersion generated from decoupling the superexchange into the particle-hole channel.

\begin{align}
    E_\parallel &= -\frac{x}{2}\epsilon_\parallel + V_\parallel (c_x + c_y) + \lambda \\
    E_\perp &= -\frac{x}{2}\epsilon_\perp+ V_{\perp}
\end{align}

$x$ here represents the doping away from half-filling (\KR{need to check my math - I think it's 2$x$ or 4$x$? It does not make a difference in any of the calculation or plots.}). The Coulomb pseudopotential $w$ is present within the dispersion. 
\end{comment}

\noindent

\bibliography{NickelateBibliography}
\end{document}